\documentclass[superscriptaddress,showpacs,preprintnumbers,amsmath,amssymb,floatfix,elsart,prd]{revtex4-2}

\usepackage{graphicx}
\usepackage{dcolumn}
\usepackage{bm}
\usepackage{epstopdf}
\usepackage{hyperref}
\hypersetup{colorlinks,citecolor=blue}

\newcommand{\beq}{\begin{equation}}
\newcommand{\eeq}{\end{equation}}
\newcommand{\bey}{\begin{eqnarray}}
\newcommand{\eey}{\end{eqnarray}}

\begin{document}

\title{Null Geodesics and QNMs in the field of Regular Black Holes }

\author{Monimala Mondal}
\email{monimala.mondal88@gmail.com} \affiliation{Department of
Mathematics, Jadavpur University, Kolkata 700032, West Bengal,
India}

\author{Anil Kumar Yadav}
\email{abanilyadav@yahoo.co.in} \affiliation{Department of Physics, United College of Engineering and Research, \\ Greater Noida, India.}

\author{Parthapratim Pradhan}
\email{pppradhan77@gmail.com} \affiliation{Department of Physics, Hiralal Mazumdar Memorial College For Women, Dakshineswar,
Kolkata-700035, India}

\author{Sayeedul Islam}
\email{sayeedul.jumath@gmail.com} \affiliation{Department of Mathematics, Amity University, \\ Kolkata , India.}

\author{Farook Rahaman}
\email{farookrahaman@gmail.com} \affiliation{Department of
Mathematics, Jadavpur University, Kolkata 700032, West Bengal,
India}

\date{\today}

\begin{abstract}
We analyze the null geodesics of regular black holes.
A detailed analysis of geodesic structure both null geodesics and time-like geodesics
have been investigated for the said black hole. As an application of null geodecics, we
calculate the radius of photon sphere and gravitational bending of light. We also study
the shadow of the black hole spacetime. Moreover, we determine the  relation between
radius of photon sphere~$(r_{ps})$ and the shadow observed by a distance observer.
Furthermore, We discus the effect of various parameters on the radius of shadow
$R_s$. Also we compute the angle of deflection for the photons as a physical
application of null-circular geodesics.  We find the relation between null geodesics
and quasinormal modes frequency in the eikonal approximation by computing the
Lyapunov exponent. It is also shown that~(in the eikonal limit) the quasinormal
modes~(QNMs) of black holes are governed by the parameter of
null-circular geodesics. The real part of QNMs frequency determines the
angular frequency whereas the imaginary part determines the instability
time scale of the circular orbit. Next we study the massless scalar perturbations
and analyze the effective potential graphically. Massive scalar perturbations also
discussed. As an application of time-like geodesics we compute the
innermost stable circular orbit~(ISCO) and marginally bound circular orbit~(MBCO)
of the regular BHs   which are closely related to the black hole
accretion disk theory. In the appendix, we calculate the relation between
angular frequency and Lyapunov exponent for null-circular geodesics.
\end{abstract}

\keywords{Photon sphere; Shadow; Quasinormal modes;  Lyaponent exponent; Regular black hole.}
\pacs{04.40.Nr, 04.20.Jb, 04.20.Dw}
\maketitle
\section{Introduction}
Geodesics motion describes the key features of black hole~(BH) space-time.
In the background spacetime, the geodesics shows a very rich
structure and  convey some important information related to the BH space-time.
There are various kinds of geodesics motions but among these, the
circular geodesics motions are more interesting due to its connection
to the gravitational binding energy. In~ \cite{Zhang/1997}, it was shown that
the binding energy of the stable circular time-like geodesics could be used to
estimate the spin of astrophysical BHs through the observations of accretion disks.
In Refs.~\cite{Kokkotas/1999,Nollert/1999,Mondal/2021}, the effectiveness of null geodesics to
explaining the characteristic modes of a BH quasinormal modes are described.
It is important to note that  general relativity~(GR) predicts precessing
elliptical orbits around a central star. In the extreme case of a Schwarzschild black
hole, there are also a simple set of unstable circular orbits which are referred as
the outcome of the non-linearity of general relativity. Cornish and Levin have
investigated that all  unstable orbits, whether regular or chaotic, can be
quantified by their Lyapunov exponents~\cite{Cornish03}.\\

Some important reviews on QNMs of astrophysical BHs and stars are given in Refs.
\cite{Nollert/1999,Kokkotas/1999,Berti/2009,Konoplya11}. In particular, it has been explained in Refs.
\cite{Nollert/1999,Kokkotas/1999} that null geodesics play an important role in describing the characteristic
modes of a BH while in Refs.~\cite{Berti/2009}, the authors have described the various aspects of
QNMs of BHs and branes. Konoplya~\cite{Konoplya11} has considered the possible
perturbations of BHs in the context of astrophysical observations.
In 1985, Mashhoon~\cite{Mashhoon/1985} has interpreted the free modes of vibrations as null particles which are
trapped at the unstable circular orbits. Later on, Berti et al.~\cite{Berti/2005} have investigated
that the null particles trapped at the unstable circular orbits are leaking out slowly. Pretorius
and Khurana~\cite{Pretorius07} have investigated that unstable circular orbits might
be useful to obtaining information on phenomena occurring at the threshold of BH formation in
the high-energy scattering of BHs. Later on Steklain and Letelier~\cite{Steklain/2009} have studied
Lyapunov exponents in the context of the stability of circular orbits for different spins
of the central body. It is well known that the unstable orbits come out with positive Lyapunov
exponents~\cite{Schnittman/2001}. Some useful applications of Lyapunov exponents have been
described in Refs.~\cite{Barrow/1981,Burd/1994,Moni19,Semerak/1999,Das/2020}. It is
important to note that the Lyapunov exponent has been known to erroneously
lead to zero Lyapunov exponents for chaotic systems therefore topological
measures of chaos are not provoked by the relativism of space and time~
\cite{Dettmann/1994,Cornish/1997}.\\

Very recent past, Prasobha and Kuriakose~\cite{Prasobha/2014} have studied
QNMs frequency of Lovelock BHs and investigated that the real part of the
modes decreases as with increase of space-time dimension. This predicts the
the presence of lower frequency modes in higher dimensions. In
Ref.~\cite{Motl/2003}, the authors have investigated QNMs of Schwarzschild
BHs in four and higher dimension in the boundary of infinite damping.
Note that the asymptotic real part of the BH QNMs holds the same frequency
as emitted by a BH whose area falls by an amount which is natural from the
point of view of distinct preludes to quantization of gravity such as
loop quantum gravity~\cite{Dreyer/2003}. Fernando and Clark \cite{Fernando/2014} have studied QNMs of
scalar perturbations in four dimensional space-time and compare their
result with Schwarzschild BH. Recently Hendi and Nemati \cite{Hendi/2020} have
investigated QNMs of scalar perturbations in five dimensional massive gravity
by using WKB method. \\

It is argued that the spacetime singularities are made due to  gravitational collapse that are always buried inside black holes.  Till now, this signature is a  major open question in general relativity. Still  we are yet to have   any healthy and consistent quantum theory of gravity that    resolves  the singularities in the interior of black holes. Therefore,   there is remarkable consideration towards the models of  regular black hole i.e. black hole  solutions without the central singularity. In Refs. \cite{Ghosh/2014,Ghosh/2015,Amir/2016}, the authors have established that the regular black holes could be contemplated  as the particle accelerator. Toshmatov  et al \cite{Toshmatov/2015} have discussed explicitly  the quasinormal modes of test fields around the regular black holes. The above investigations  motivate us   to study the regular black holes. The main focus of the work is to investigate the geodesics structure of
regular black holes. We have studied both null geodesics as well
as time-like geodesics. Aa an application of null geodesics, we have derived
the radius of photon sphere and study the shadow of the BH visually. Also
we find the relation between radius of photon sphere~$(r_{ps})$ and the shadow as
observed by a distance observer. Furthermore, we discus the implication of various parameters on the
radius of shadow $R_s$. Also we derive the angle of deflection for
the photons.  By computing the Lyapunov exponent we find the relation
between null geodesics and QNMs frequency in the eikonal approximation.
Also we study the massless scalar perturbations and analyze the effective
potential graphically.  Moreover we compute the ISCO and MBCO  of the said  which are closely related to the BH
accretion disk theory.\\

The paper is organized as follows. In the next section, we briefly discuss the regular
BHs. In Sec.~III, we investigate thoroughly the null geodesics of the said BH.  In Sec.~IV, we study the Klien-Gordon equations for massless perturbations and derived the effective
potential. We plotted it graphically and compared the result with Hayward class and Bardeen class of BHs. Sec.~V, is described to study the massive scalar perturbations and computed the effective potential. In Sec.~VI, we study the relation between QNMs frequency and null circular
geodesics through massless scalar perturbations in the eikonal limit. In Sec.~VII, we investigated the time-like geodesics in the said BH. We also compute the ISCO and MBCO. In Sec.~VIII, we have given the conclusions and future outlook.

\section{Introduction of Regular BHs  }
In this section, we will consider a static, spherically symmetric and
asymptotically flat solutions of regular BH  ~\cite{Fan/2016} as
follows:
\begin{eqnarray}
\label{met}
ds^{2} = -f(r) dt^{2} + \frac{dr^{2}}{f(r)} + r^{2} (d\theta^{2} + \sin^{2}\theta d\phi^{2}),
\end{eqnarray}
where
\begin{eqnarray}
\label{fr}
f(r) =  \bigg(1-\frac{2 M}{r} - \frac{2 \alpha^{-1}q^{3} r^{\mu-1}}{(r^\nu +q^\nu)^\frac{\mu}{\nu}} \bigg)
\end{eqnarray}
\begin{figure}[h!t]
\centering
\includegraphics[width=7cm,height=5cm,angle=0]{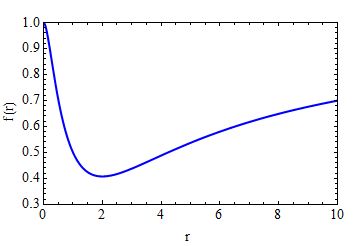}
\includegraphics[width=7cm,height=6cm,angle=0]{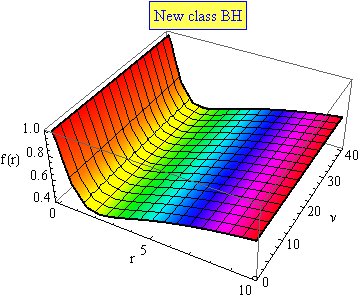}
\includegraphics[width=7cm,height=5cm,angle=0]{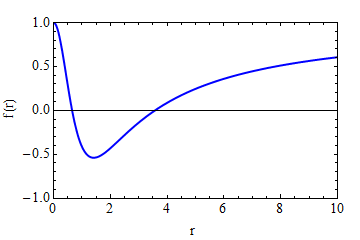}
\includegraphics[width=7cm,height=6cm,angle=0]{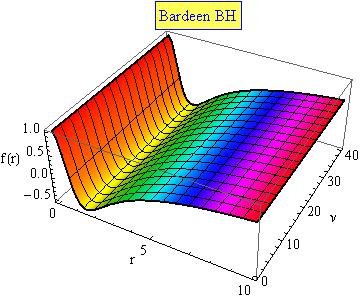}
\includegraphics[width=7cm,height=5cm,angle=0]{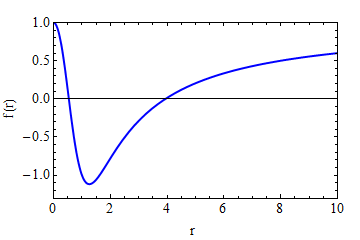}
\includegraphics[width=7cm,height=6cm,angle=0]{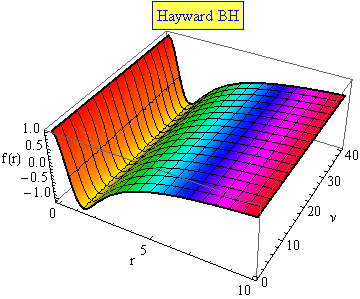}
\caption{$f(r)$ versus $r$ with $\nu=1$ for New class BH~(upper panel),  $\nu=2$ for
Bardeen BH~(middle panel)and $\nu=3$ for Hayward BH~(lower panel).
Here, we set $M = 0$, $q = 1$, $\mu = 3$ and $\alpha = 0.5$.
The right panel is plotted for values of $\nu$ in the
range $0\leq \nu \leq 40$.}
\end{figure}
\begin{figure}[h!]
\centering
\includegraphics[width=8cm,height=6cm,angle=0]{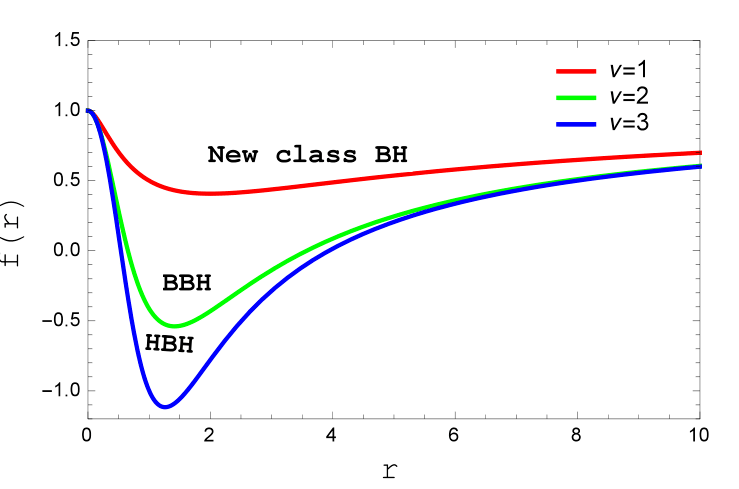}
\caption{$f(r)$ versus $r$ for  various values of $\nu$;  the
other parameters fixed to $M = 0$, $q = 1$, $\mu = 3$ and $\alpha = 0.5$.}
\end{figure}
Here, $\mu > 0 $ is a dimensionless constant, $\alpha > 0 $ has the
dimension of length squared and $ q $ is a free integrating constant.\\

Special case:\\
(A). Bardeen class:\\
If $\nu =2$, then the function (\ref{fr}) is reduced to
\begin{eqnarray}
f = \bigg(1-\frac{2 M}{r} - \frac{2 \alpha^{-1}q^{3} r^{\mu-1}}{(r^2 +q^2)^\frac{\mu}{2}} \bigg).
\end{eqnarray}
For $ M = 0, \mu = 3$, the solution is the Bardeen BH (BBH)~\cite{Bardeen74}.\\
(B). Hayward class:\\
If $\nu = \mu$,  then the function (\ref{fr}) is reduced to
\begin{eqnarray}
f = \bigg(1-\frac{2 M}{r} - \frac{2 \alpha^{-1}q^{3} r^{\mu-1}}{(r^\mu +q^\mu)} \bigg).
\end{eqnarray}
For $ M = 0, \mu = 3$, the solution is the Hayward BH (HBH)~\cite{Hayward06}.\\
(C). A new class:\\
If $\nu =1$, then the function (\ref{fr}) is reduced to
\begin{eqnarray}
f = \bigg(1-\frac{2 M}{r} - \frac{2 \alpha^{-1}q^{3} r^{\mu-1}}{(r +q)^\mu} \bigg).
\end{eqnarray}
For $ M = 0$, the solution can be treated as a new class solution. Also, regular
BH has a solution for $ \mu \geq 3$ .\\
The variation of $f(r)$ versus  $r$ could be seen from figures 1 and 2. From the figures we can see that, when $\nu = 1$ the solution New class BH has no zero, when $\nu = 2,3$ the solution BBH and HBH both has two horizon, respectively.

\section{ Action and field equations}
The action for Einstein gravity coupled to a non-linear electromagnetic field is read as
\begin{equation}
\label{A-1}
I = \frac{1}{16\pi}\int\sqrt{-g}\left(R-\mathcal{L}(\mathcal{F})\right)d^{4}x
\end{equation}
where $\mathcal{F} \equiv F_{ij}F^{ij}$ with $F = dA$ as field strength of the vector field and Lagrangian
density $\mathcal{L}$ is a function of $\mathcal{F}$.\\
Also, the co-variant equation of motions are given by
\begin{equation}
\label{A-2}
G_{ij} = T_{ij},\;\;\;\triangledown_{i}\left(\mathcal{L}_{\mathcal{F}}F^{ij}\right) = 0
\end{equation}
where $\mathcal{L}_{\mathcal{F}} = \frac{\partial \mathcal{L}}{\partial \mathcal{F}}$ and $G_{ij} = R_{ij}-\frac{1}{2}Rg_{ij}$. $R$ and $g$ have their usual meaning as in Einstein tensor.\\
The energy momentum tensor $T_{ij}$ in this case is defined as
\begin{equation}
\label{A-3}
T_{ij} = 2\left(\mathcal{L}_{\mathcal{F}}F_{ij}^{2}-\frac{1}{4}g_{ij}\mathcal{L}\right)
\end{equation}
In this paper, we have considered a static, spherically symmetric and asymptotically flat solutions of regular BH.
\begin{eqnarray}
\label{A-4}
ds^{2} = -f(r) dt^{2} + \frac{dr^{2}}{f(r)} + r^{2} (d\theta^{2} + \sin^{2}\theta d\phi^{2}),
\end{eqnarray}
Thus, the independent field equations are read as
\begin{equation}
\label{A-5}
0 = \frac{f^{\prime}}{r}+\frac{f-1}{r^{2}}+\frac{1}{2}\mathcal{L}
\end{equation}
\begin{equation}
\label{A-6}
0 = f^{\prime\prime}+\frac{2f^{\prime}}{r}+\mathcal{L}-\frac{q^{8}}{\alpha^{2}r^{4}}\mathcal{L}_{\mathcal{F}}
\end{equation}
A detail description of action and field equations of regular BH   are found in Ref. ~\cite{Fan/2016}.   \\
Here, we have assumed three special class\\
(A). Bardeen class with Lagrangian density $\mathcal{L} = \frac{4\mu}{\alpha}\frac{(\alpha\mathcal{F})^{\frac{5}{4}}}{\left(1+\sqrt{\alpha \mathcal{F}}\right)^{1+\frac{\mu}{2}}}$\\
(B). Hayward class with Lagrangian density $\mathcal{L} = \frac{4\mu}{\alpha}\frac{(\alpha\mathcal{F})^{\frac{\mu+3}{4}}}{\left(1+(\alpha \mathcal{F})^{\frac{\mu}{4}}\right)^{2}}$ \\
(C). New class with Lagrangian density $\mathcal{L} = \frac{4\mu}{\alpha}\frac{\alpha\mathcal{F}}{\left(1+(\alpha \mathcal{F})^{\frac{1}{4}}\right)^{\mu+1}}$\\
\subsection{Circular orbits in the Equatorial Plane}
In an equatorial plane, we compute the geodesic for the space-time (\ref{met}) by using
the method of Chandrasekhar et al. \cite{Chandrasekhar83}. In order to do that, we
consider $\dot{\theta} = 0$ and   $\theta$ = constant = $\frac{\pi}{2}$. Now, the
Lagrangian equation of motion becomes
\begin{eqnarray}
2\mathcal{L} = \bigg[- \bigg(1-\frac{2 M}{r} - \frac{2 \alpha^{-1}q^{3} r^{\mu-1}}{(r^\nu +q^\nu)^\frac{\mu}{\nu}} \bigg) \dot{t}^{2} + \frac{\dot{r}^{2}}{\bigg(1-\frac{2 M}{r} - \frac{2 \alpha^{-1}q^{3} r^{\mu-1}}{(r^\nu +q^\nu)^\frac{\mu}{\nu}} \bigg)} + r^{2} \dot{\phi}^{2} \bigg],
\end{eqnarray}
where $\phi$ denotes the angular momentum. Now the canonical momentum is defined as
\begin{eqnarray}
\label{momenta}
P_q = \frac{\partial \mathcal{L} }{\partial \dot{q}} .
\end{eqnarray}
By using the equation (\ref{momenta}), the generalized momenta can be derived from the lagrangian are as follows
\begin{eqnarray}
\label{pot}
p_{t}& =& -\bigg(1-\frac{2 M}{r} - \frac{2 \alpha^{-1}q^{3} r^{\mu-1}}{(r^\nu +q^\nu)^\frac{\mu}{\nu}} \bigg) \dot{t}
= - E = \textrm{const}. \\
\label{pof}
p_{\phi}& =& r^{2} \dot{\phi} = L = \textrm{const}. \\
\label{por}
p_{r} &= & \frac{\dot{r}}{ \bigg(1-\frac{2 M}{r} - \frac{2 \alpha^{-1}q^{3} r^{\mu-1}}{(r^\nu +q^\nu)^\frac{\mu}{\nu}} \bigg)}.
\end{eqnarray}
Here, the Lagrangian equation of motion is not depends on $`t'$ and $`\phi'$ both, thus $p_{t}$ and $p_{\phi}$ are
conserved quantities. After solving equations (\ref{pot}) and (\ref{pof}) for $\dot{t}$ and $\dot{\phi}$, we get
\begin{eqnarray}
\label{dot}
\dot{t} = \frac{E}{\bigg(1-\frac{2 M}{r} - \frac{2 \alpha^{-1}q^{3} r^{\mu-1}}{(r^\nu +q^\nu)^\frac{\mu}{\nu}} \bigg)}~~~~ \textrm{and} ~~~~\dot{\phi} = \frac{L}{r^{2}}.
\end{eqnarray}
Integral equation of the geodesic motion is obtained by
normalizing the four velocity $(u^{a})$ as follows
\begin{eqnarray}
g_{a b} u^{a}u^{b} = \delta,
\end{eqnarray}
which is similar to
\begin{eqnarray}
\label{jai.}
-E \dot{t} + L \dot{\phi} + \frac{\dot{r}^{2}}{\bigg(1-\frac{2 M}{r}
- \frac{2 \alpha^{-1}q^{3} r^{\mu-1}}{(r^\nu +q^\nu)^\frac{\mu}{\nu}} \bigg)}
=  \delta.
\end{eqnarray}
Here, $\delta = -1$ determines the time-like geodesic, $\delta = 0$ determines the null
geodesic and $ \delta = 1 $ determines the space-like geodesic. Putting the value of
$\dot{t}$ and $\dot{\phi}$  from (\ref{dot}) into (\ref{jai.}), we find the radial
equation  for space-time as
\begin{eqnarray}
\label{dotr}
\dot{r}^{2} = E^{2} - \bigg( \frac{L^{2}}{r^{2}} -\delta\bigg)\bigg(1-\frac{2 M}{r}
- \frac{2 \alpha^{-1}q^{3} r^{\mu-1}}{(r^\nu +q^\nu)^\frac{\mu}{\nu}} \bigg).
\end{eqnarray}

\section{Null Geodesics of Regular BHs   }
The radial equation of the test particle for null circular geodesic as using the
equation~(\ref{dotr}) by setting $\delta = 0$ is given by
\begin{eqnarray}
\label{radeqn}
\dot{r}^{2} = E^{2} -V_{null}=E^{2}- \frac{L^{2}}{r^{2}}\bigg(1-\frac{2 M}{r}
- \frac{2 \alpha^{-1}q^{3} r^{\mu-1}}{(r^\nu +q^\nu)^\frac{\mu}{\nu}} \bigg).
\end{eqnarray}
where $V_{null}$ is the effective potential of null-circular geodesics is given by
\begin{eqnarray}
V_{null} = \frac{L^{2}}{r^{2}}\bigg(1-\frac{2 M}{r} -
\frac{2 \alpha^{-1}q^{3} r^{\mu-1}}{(r^\nu +q^\nu)^\frac{\mu}{\nu}} .
\end{eqnarray}

\subsection{Radial null geodesics:}
The radial geodesics is corresponding to zero angular momentum~(L=0).
Hence the effective potential for radial null geodesics is
\begin{eqnarray}
V_{null} =0
\end{eqnarray}
The equation for $\dot{t}$ and $\dot{r}$ are simplified to
\begin{eqnarray}
\dot{r} =\pm E~~~~~~~~ \textrm{and}~~~~~~~~ \dot{t}= \frac{E}{f(r)}.
\end{eqnarray}
The above equation gives
\begin{eqnarray}
\frac{d t}{d r} = \pm \frac{1}{f(r)} = \pm \frac{1}{\bigg(1-\frac{2 M}{r}
- \frac{2 \alpha^{-1}q^{3} r^{\mu-1}}{(r^\nu +q^\nu)^\frac{\mu}{\nu}} \bigg)}.
\end{eqnarray}
The above equation can be integrated to find the coordinate time $t$ as
\begin{eqnarray}
t =  \pm \int \frac{1}{\bigg(1-\frac{2 M}{r} -
\frac{2 \alpha^{-1}q^{3} r^{\mu-1}}{(r^\nu +q^\nu)^\frac{\mu}{\nu}} \bigg)}+ \textrm{constant}.
\end{eqnarray}
When $ r\rightarrow 2M$ and $q\rightarrow0$, $t\rightarrow\infty$.
Hence we can obtain the proper time by integrating
\begin{eqnarray}
\frac{d \tau}{d r} = \pm \frac{1}{E},
\end{eqnarray}
which gives
\begin{eqnarray}
\tau = \pm \frac{r}{E}+ \textrm{constant}.
\end{eqnarray}
When $ r\rightarrow 2M$ and $q\rightarrow0$, $\tau \rightarrow  \pm \frac{2M}{E}$
is finite. Hence the coordinate time is infinite while the proper time is finite.
This is the same with the result for the Schwarzschild BH.

\subsection{Geodesics with angular momentum$(L\neq 0)$:}
\begin{figure}
\begin{center}
\includegraphics[width=6cm,height=3.5cm,angle=0]{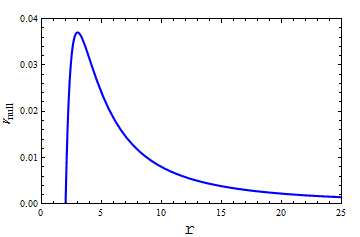}
\includegraphics[width=6cm,height=4.5cm,angle=0]{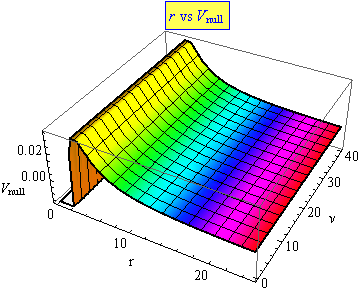}
\includegraphics[width=6cm,height=3.5cm,angle=0]{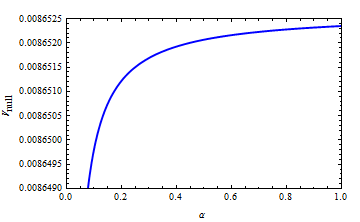}
\includegraphics[width=6cm,height=4.5cm,angle=0]{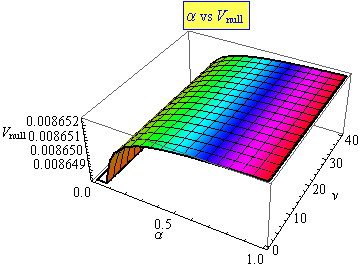}
\includegraphics[width=6cm,height=3.5cm,angle=0]{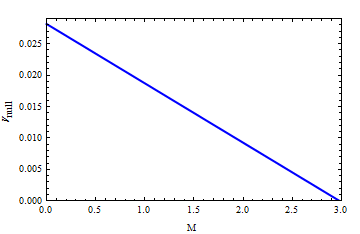}
\includegraphics[width=6cm,height=4.5cm,angle=0]{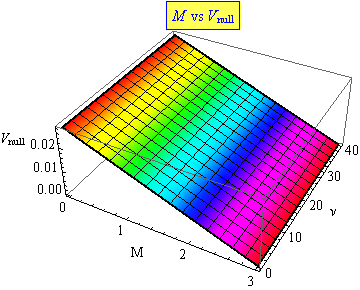}
\includegraphics[width=6cm,height=3.5cm,angle=0]{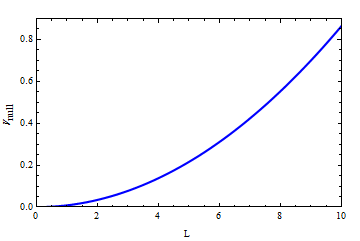}
\includegraphics[width=6cm,height=4.5cm,angle=0]{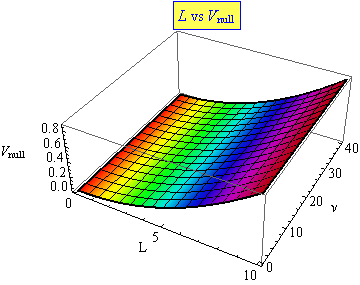}
\includegraphics[width=6cm,height=3.5cm,angle=0]{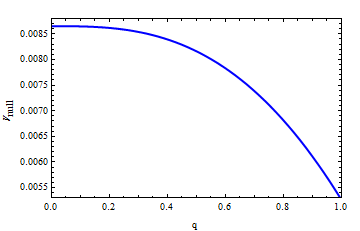}
\includegraphics[width=6cm,height=4.5cm,angle=0]{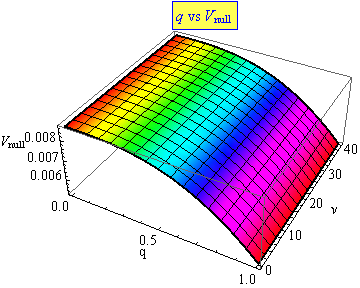}
\end{center}
\caption{The figure describes the variation  of $V_{null}$ with $r$~(first panel), $M$~(second panel), $q$~(third panel), $\alpha$~(fourth panel)
and $L$~(fifth panel). We have set ${\mu}=3$.
\label{vnull}}
\end{figure}

In this case, the effective potential is
\begin{eqnarray}
V_{null} = \frac{L^{2}}{r^{2}}\bigg(1-\frac{2 M}{r} -
\frac{2 \alpha^{-1}q^{3} r^{\mu-1}}{(r^\nu +q^\nu)^\frac{\mu}{\nu}} \bigg).
\end{eqnarray}
We have plotted the graph $V_{null}$ against $r$~(first panel) in the Fig.~\ref{vnull}.
Initially, $V_{null}$  reaches a peak value then it  decreases for larger value of $r$
as evident from Fig~\ref{vnull}. In the same figure one can see that $V_{null}$ is
linearly decreasing with $M$~(second panel), then it decreases with decreasing value of $q$~(third panel)
while $V_{null}$ increases with increasing value of $\alpha$~(fourth panel)
and $L$(fifth panel) respectively.\\
In case of circular geodesics ~\cite{Chandrasekhar83}
\begin{eqnarray}
\label{deri}
\dot{r}^{2} = (\dot{r}^{2})' = 0,
\end{eqnarray}
Now, the angular momentum and energy at $r = r_o$ for the null geodesics are as follows:
\begin{eqnarray}
\label{imp}
\bigg(r^\mu_{o} q^3((\mu - 3 ) q^\nu - 3 r^\nu_{o}) + ( r_{o} - 3M) {(r^\nu_{o} +q^\nu)^\frac{\mu +\nu}{\nu}} \alpha\bigg) = 0 ~~~\textrm{and}\nonumber\\
\frac{E_{o}}{L_{o}} = \pm \sqrt{\frac{\bigg((r- 2 M)(r^\nu_{o} +q^\nu)^\frac{\mu}{\nu}-2 \alpha^{-1}q^3 r^\mu_{o} \bigg)}{r_{o}^{2}(r^\nu_{o} +q^\nu)^\frac{\mu}{\nu}}}.
\end{eqnarray}
Let us consider $ D_{o} = \frac{L_{o}}{E_{o}}$ be the impact parameter, then the equation (\ref{imp}) reduces  to
\begin{eqnarray}
\frac{1}{ D_{o}} = \frac{E_{o}}{L_{o}} = \sqrt{\frac{M +q^3 r^\mu_{o} q^3 \alpha^{-1} (r^\nu_{o} + (1- \mu)q^\nu){(r^\nu_{o} +q^\nu)^{(-\frac{\mu +\nu}{\nu})}} }{r^3_{o}}} .
\end{eqnarray}

\subsection{Radius of photon sphere :}
A photon sphere is a area where the gravitational field of BH  is so strong that light can travel in
circles. At the photon sphere, no light released outside can reach observer from below- the observer
watch into the vast wide emptiness of the BH.
For spherical geodesics of a circular light orbits, we use two conditions\\
~~~~~~~~~~~~~~$V_{null}(r) |_{r_o = r_{ps}} =0$~~~ and ~~~~$\frac{\partial V_{null}(r)}{\partial r} |_{r_o = r_{ps}} =0$,\\
where, $r_{ps}$ is the radius of the photon orbit.
The first condition gives
\begin{eqnarray}
\frac{L^2}{E^2} = \frac{r_{ps}^2}{f(r_{ps})},
\end{eqnarray}
and the second condition implies
\begin{eqnarray}
r_{ps}f'(r_{ps}) - 2 f(r_{ps}) = 0.
\end{eqnarray}
Substituting the equation (\ref{fr}) into the above  equation, we have
\begin{eqnarray}
\label{eqnps}
r^\mu_{ps} q^3((\mu - 3 ) q^\nu - 3 r^\nu_{ps}) + ( r_{ps} - 3M) {(r^\nu_{ps} +q^\nu)^\frac{\mu +\nu}{\nu}} \alpha = 0
\end{eqnarray}
The analytical solution $r_{ps}$ of the above equation is not a trivial task.
But for some special cases, we can find a relation for the radius of the photon sphere, as\\
\begin{figure*}
\begin{tabular}{rl}
\includegraphics[width=0.35\textwidth]{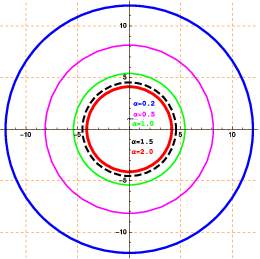}~~~~~~~~~~~~~~~~~~
\includegraphics[width=0.35\textwidth]{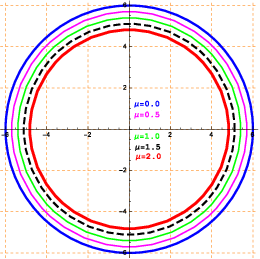}\\\\\\
\includegraphics[width=0.35\textwidth]{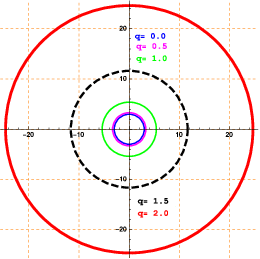}~~~~~~~~~~~~~~~~~~
\includegraphics[width=0.35\textwidth]{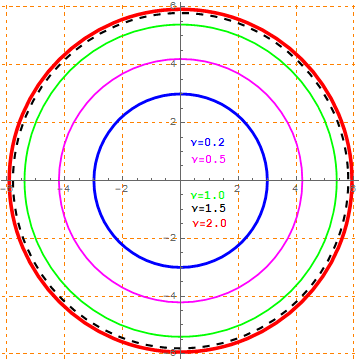}\\
\end{tabular}
\caption{Black hole shadow in the Celestial plane $\beta-\gamma$ for
varying $\alpha$ with $\mu=1, \nu=1$ and $q = 1$~(up- left panel), varying $\mu$ with
$\alpha=1, \nu=1$ and $q = 1$~(up- right panel), varying $q$ with  $\alpha=1, \mu=1$ and
$\nu= 1$~(down left panel), varying $\nu$ with  $\alpha=1, \mu=1$ and $q = 1$~(down- left panel).
We set $M = 1$.}
\label{6amnq}
\end{figure*}
$\bullet$ when $\alpha = 0$, the radius of photon sphere is follows
\begin{eqnarray}
r_{ph}= 3^{-\frac{1}{\nu}} q^{-\frac{3}{\nu}} \bigg(q^{3+\nu}(-3-\mu)\bigg)^{\frac{1}{\nu}}
\end{eqnarray}
$\bullet \bullet$ when $q = 0$,  the radius of the photon sphere is
\begin{eqnarray}
r_{ps} = 3M
\end{eqnarray}
$\bullet \bullet \bullet$ when $\mu = 0$, the radius of the photon sphere
is as follows
\begin{eqnarray}
r_{ps}= 3(-q^{\nu})^{\frac{1}{\nu}}(M + \frac{q^3}{\alpha})
\end{eqnarray}
$\bullet \bullet \bullet \bullet$ when $\nu =1(\nu \neq 0) $, the relation for
the radius of the photon sphere is as follows
\begin{eqnarray}
r^\mu_{ps} q^3((\mu - 3 ) q - 3 r_{ps}) + ( r_{ps} - 3M) (r_{ps} +q)^{\mu} \alpha = 0
\end{eqnarray}
\begin{table}
\begin{center}
\caption{Photon sphere radius for variation of massive  parameters with $M=1$.}\label{tab1}
\begin{tabular}{|p{3.5cm} p{1.5cm} p{1.5cm} p{1.5cm} p{1.5cm} l|}
\hline
$\alpha$ &0.2& 0.5 & 1.0 & 1.5 & 2.0 \\
\hline
$r_{ps}(\mu=\nu=q=1)$& 16.99&8.15& $5.4$ & $4.54$ &4.12 \\
\hline
\hline
$\mu$ &0.0& 0.5 & 1.0 &1.5& 2.0\\
\hline
$r_{ps}(\alpha=\nu=q=1)$ &6.0 & 5.7& 5.4& 5.11&4.82 \\
\hline
\hline
$\nu$ &0.2& 0.5 & 1.0 & 1.5 &2.0 \\
\hline
$r_{ps}(\mu=\alpha=q=1)$ &3.14& 4.21 &5.4 &5.8&5.93\\
\hline \hline
$q$ &0.0& 0.5 & 1.0 &1.5& 2.0\\
\hline
$r_{ps}(\mu=\nu=\alpha=1)$ &3.0& 3.31 & 5.4 &11.63&24.64\\
\hline
\end{tabular}
\end{center}
\end{table}
In general, it is possible to numerical that there are many
roots~(real, complex according to the value of the parameter $\alpha, \mu, \nu$ and $q$) of the
Eq.~(\ref{eqnps}), which are bigger than the event horizon radius and then we have many small and
larger spherical light orbits. To understand which one is stable with respect to radial perturbation,
we should examine the sign of $V''_{null}$, when $V''_{null} < 0$ it indicates unstable orbit and
$V''_{null} > 0$  indicates the stable one.
In the following Table~\ref{tab1}, the radius of the unstable photon sphere~(the larger photon sphere)
is listed for some parameters, thereby we conclude the nature of $r_{ps}$ under  variation of the
massive parameters.

We plot the BH shadow in Fig.~\ref{6amnq} for different values of parameters. We observe that in both
cases $\alpha$ and $\mu$ when the values increase the radius decrease while keeping other parameter
fixed. We observe that the shadow size shrinks with increasing $`` \alpha"$ and $``\mu"$ , respectively.
Further, we notice that with increase in $``q"$, the shadow size increase with fixed other parameters.
The shadow size also increase with increase in $``\nu"$. One can observed from Figure \ref{6amnq} that
$\mu$ has weaker effect whereas $q$ has stronger effect on the BH shadow.

\subsection{Shadow of the BH:}
Now, we study the shadow Ref.\cite{Kimet/20} of the BH. We assume that a bright object released photons and after
releasing from the bright object, photons comes towards the BH which is situated between a
bright object and an observer. Around the BH there are three possible trajectories of the
photon geodesics: (i) falling into the BH, (ii) scattered away from BH to infinity,
(iii) first two sets are separated by critical geodesics. The observer can see only
the scattered photons which fall from a dark region into the BH.
This dark region is called BH shadow.\\

Now, to study the shadow of the BH we are going to introduce a new celestial
coordinate $(\beta,\gamma)$, where $\beta$ is the  perpendicular distance of
the shadow from symmetry axis and $\gamma$ is the apparent perpendicular
distance of the shadow from its projection on the equatorial plane.
Following the calculation of Ref.\cite{Das20} we can obtain an equation
representing a circle of radius in celestial plane $\beta-\gamma$, as follows
\begin{eqnarray}
\beta^2 + \gamma^2 = R_s ^2= \frac{\frac{r_{ps}^2}{f(r_{ps})}}{1-\frac{r_{ps}}{f(r_{ps})}.\frac{f(r_{ob})}{r_{ob}^2}},
\end{eqnarray}
where subscripts $``ps"$ and $``ob"$ represents the photon sphere and observer, respectively.
Considering the equation (\ref{fr}), for distant observer when $r\rightarrow \infty$, we can
find that $\frac{f(r_{ob})}{r^2_{ob}}\rightarrow 0$. So the radius of the shadow($R_s$) reduces
to
\begin{eqnarray}
\label{radsha}
R_s = \frac{r_{ps}}{\sqrt{f(r_{ps})}}
\end{eqnarray}
\begin{figure*}
\begin{tabular}{rl}
\includegraphics[width=0.45\textwidth]{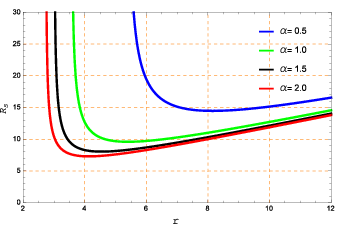}~~~~~~~~~~~~~~~~~~~~
\includegraphics[width=0.33\textwidth]{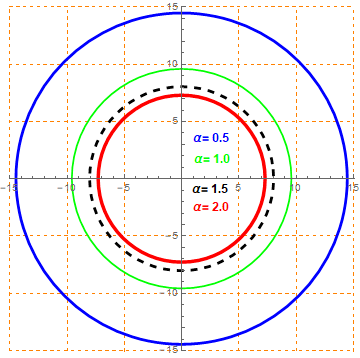}\\
\includegraphics[width=0.45\textwidth]{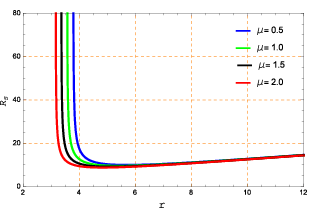}~~~~~~~~~~~~~~~~~~~~
\includegraphics[width=0.33\textwidth]{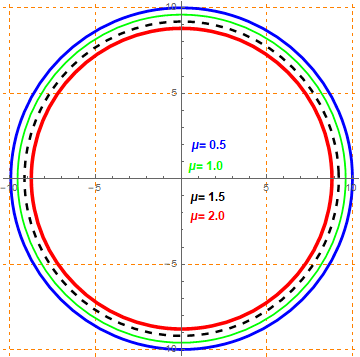}\\
\end{tabular}
\caption{The figure shows $R_s$ versus $r$ (up-left Panel) for varying $ \alpha$ and varying $ \mu$ (down-left Panel). BH shadow in the Celestial plane $\beta-\gamma$ for varying $\alpha$ with $\mu=1, \nu=1$ and $q = 1$(up- left panel), varying $\mu$ with $\alpha=1, \nu=1$ and $q = 1$(down- right panel). We set $M = 1$. }
\label{7almu}
\end{figure*}
\begin{figure*}
\begin{tabular}{rl}
\includegraphics[width=0.43\textwidth]{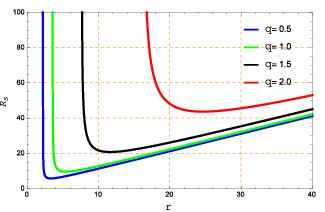}~~~~~~~~~~~~~~~~~~~~
\includegraphics[width=0.33\textwidth]{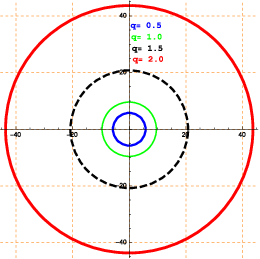}\\
\includegraphics[width=0.43\textwidth]{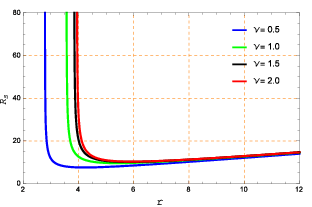}~~~~~~~~~~~~~~~~~~~~
\includegraphics[width=0.33\textwidth]{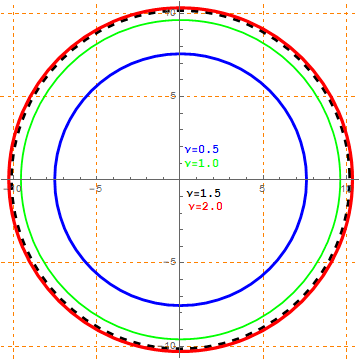}\\
\end{tabular}
\caption{The figure shows $R_s$ versus $r$ (up-left Panel) for varying $ q$ and varying $ \nu$ (down-left Panel). Black hole shadow in the Celestial plane $\beta-\gamma$ for varying $q$ with $\alpha=1, \mu=1$ and $\nu$(up- right panel), varying $\nu$ with  $\alpha=1, \mu=1$ and $q = 1$(down- right panel). We set the parameter $M = 1$.}
\label{7qnu}
\end{figure*}
\subsection{Dependence of shadow radius $R_s$ on various parameters:}
From Eq.~(\ref{radsha}) we can write the expression of the shadow radius $R_s$ as
follows
\begin{eqnarray}
R_s = \frac{r_{ps}}{\sqrt{ \bigg(1-\frac{2 M}{r_{ps}} -
\frac{2 \alpha^{-1}q^{3} r_{ps}^{\mu-1}}{(r_{ps}^\nu +q^\nu)^\frac{\mu}{\nu}} \bigg)}}
\end{eqnarray}
The above expression of $R_s$ shows the effects of the parameter $\alpha, \mu, \nu$ and $q$
on the silhouette of the shadow. The variation in the silhouette of the BH shadow are
shown graphically  in Figs.~$\ref{7almu}$, $\ref{7qnu}$ for different values of parameters.

In Table~\ref{tab2} we represent the computed  the values of radius of photon sphere~($r_{ps}$)
and the BH shadow radius $(R_s)$ for different values of the parameters.

In Fig.~\ref{7almu}, $R_s$ is plotted against  $r$  upper and lower left panel for varying
$\alpha$ and $\mu$, respectively and the  BH shadow against $R_s$ is plotted upper and
lower right panel for  varying $\alpha$ and $\nu$, respectively. From Figure~($\ref{7almu}$)
we observe that the shadow size shrinks with increasing $`` \alpha"$ and $``\mu"$  when  the
value of  the other parameters is fixed.

In Fig.~\ref{7qnu}, $R_s$ is plotted against  $r$  upper and lower left panel for varying $q$
and $\nu$, respectively. The  BH shadow against $R_s$ is plotted upper and lower right panel
for  varying $q$ and $\mu$, respectively. One can see from Figure~($\ref{7qnu}$) that the shadow
size increase with increasing $``q"$ and $``\nu"$, the shadow radius $R_s$ increase , while other
parameters is fixed.\\
Also one can find that variation of $\mu$ has weaker effect on the shadow size than the other
parameter and $q$ has more significant effect on the BH shadow.
\subsection{Gravitional Bending of light}
A  non-stable circular photon orbit is called ``Photon Sphere". This non-stable photon sphere
represents the shadow of the BH.
From Eq.~(\ref{pof}) and Eq.~(\ref{por}) we have
\begin{eqnarray}
\label{drphi}
\frac{d r }{d \phi} = \frac{\dot{r}}{\dot{\phi}} = \frac{p_r   \bigg(1-\frac{2 M}{r} -
\frac{2 \alpha^{-1}q^{3} r^{\mu-1}}{(r^\nu +q^\nu)^\frac{\mu}{\nu}} \bigg) r^2}{L}.
\end{eqnarray}
Again the Eq.~(\ref{radeqn}) can be rewritten as
\begin{eqnarray}
\label{pr2}
 p_r^2  \bigg(1-\frac{2 M}{r} - \frac{2 \alpha^{-1}q^{3} r^{\mu-1}}{(r^\nu +q^\nu)^\frac{\mu}{\nu}} \bigg)
 =\frac{ E^2}{  \bigg(1-\frac{2 M}{r} - \frac{2 \alpha^{-1}q^{3} r^{\mu-1}}{(r^\nu +q^\nu)^\frac{\mu}{\nu}} \bigg)}
 - \frac{L^2}{r^2} .
\end{eqnarray}
We can get $p_r$ from the Eq.~(\ref{pr2}) as
\begin{eqnarray}
\label{pr}
p_r = \pm  \sqrt{\frac{1}{\bigg(1-\frac{2 M}{r}
- \frac{2 \alpha^{-1}q^{3} r^{\mu-1}}{(r^\nu +q^\nu)^\frac{\mu}{\nu}} \bigg)}}
\sqrt{\frac{ E^2}{\bigg(1-\frac{2 M}{r} -
\frac{2 \alpha^{-1}q^{3} r^{\mu-1}}{(r^\nu +q^\nu)^\frac{\mu}{\nu}} \bigg)} - \frac{L^2}{r^2}}.
\end{eqnarray}
Using $p_r$ from the above equation, we can write the Eq.~(\ref{drphi}) as follows
\begin{eqnarray}
\label{ndrph}
\frac{d r }{d \phi} = \pm \sqrt{\bigg(1-\frac{2 M}{r} -
\frac{2 \alpha^{-1}q^{3} r^{\mu-1}}{(r^\nu +q^\nu)^\frac{\mu}{\nu}} \bigg)}
\sqrt{\frac{E^2 }{L^2}\chi^2(r) - 1 },
\end{eqnarray}
where,
\begin{eqnarray}
\chi^2(r) = \frac{r^2}{\bigg(1-\frac{2 M}{r} - \frac{2 \alpha^{-1}q^{3} r^{\mu-1}}{(r^\nu +q^\nu)^\frac{\mu}{\nu}} \bigg)}
\end{eqnarray}
Now, a light ray coming from infinity, reaches at minimum radius ~($R$)
and again returns back to infinity, the bending angle~$(\beta_{bending})$ is
read by the formula
\begin{eqnarray}
\beta_{bending} = -\pi +2 \int_R ^\infty \frac{dr}
{\sqrt{r^2  \bigg(1-\frac{2 M}{r} - \frac{2 \alpha^{-1}q^{3} r^{\mu-1}}{(r^\nu +q^\nu)^\frac{\mu}{\nu}} \bigg)\bigg(\frac{E^2 }{L^2}\chi^2(r)-1 \bigg)}}
\end{eqnarray}
As $R$ is the turning point of the trajectory, the  necessary condition $\frac{d r }{d \phi}|_{R = 0}$
must be satisfied.

This leads to the equation
\begin{eqnarray}
\chi^2(R) = \frac{L^2 }{E^2}.
\end{eqnarray}
Then the deflection angle as a function of $R$ can be written as
\begin{eqnarray}
\beta_{bending} = -\pi +2 \int_R ^\infty \frac{dr}
{\sqrt{r^2  \bigg(1-\frac{2 M}{r} - \frac{2 \alpha^{-1}q^{3} r^{\mu-1}}{(r^\nu +q^\nu)^\frac{\mu}{\nu}} \bigg)\bigg(\frac{\chi^2(r)}{\chi^2(R)}-1 \bigg)}}
\end{eqnarray}
Substituting  the value of $\chi^2(r)$ and $\chi^2(R)$, the equation of bending angle of regular BH  is
\begin{eqnarray}
\beta_{bending} = -\pi +
2 \int_R ^\infty \frac{dr}
{\sqrt{r^2  \bigg(1-\frac{2 M}{r} - \frac{2 \alpha^{-1}q^{3} r^{\mu-1}}{(r^\nu +q^\nu)^\frac{\mu}{\nu}} \bigg)
\bigg(\frac{r^2}{D^2\left(1-\frac{2 m r^{2}}{r^{3} + 2l^{2} m}\right)}-1 \bigg)}}
\end{eqnarray}
where, $D = \frac{L}{E}$ is the impact parameter of the regular BH.
The exact formula of bending angle is derived in \cite{chiba17} .

\begin{table*}
\small
\caption{Radius of photon sphere $r_{ps}$ of the BH, shadow radius $R_s$
for variation of massive  parameters with $M=1$. } \label{tab2}
\begin{tabular}{p{1.5cm} p{1.5cm}  p{3.5cm} p{3.5cm} }\\
\hline
S. N. & $\alpha$ & $r_{ps}(\mu=\nu=q=1)$ & $R_s(\mu=\nu=q=1)$ \\
\hline
1. & 0.5 &  $8.15$ & $14.47$   \\
2. & 1.0 &  $5.4$  & $9.59$   \\
3. & 1.5 &  $4.54$ & $8.04$   \\
4. & 2.0 &  $4.12$ & $7.29$  \\
\hline
\hline
S. N. & $\mu$ & $r_{ps}(\alpha=\nu=q=1)$ & $R_s(\mu=\nu=q=1)$ \\
\hline
1. & 0.5 & $5.7$ & $9.99$   \\
2. & 1.0 &  $5.4$ & $9.59$   \\
3. & 1.5 & $5.11$ & $9.19$   \\
4. & 2.0 &  $4.82$ & $8.79$  \\
\hline
\hline
S. N. & $\nu$ & $r_{ps}(\alpha=\mu=q=1)$ & $R_s(\mu=\nu=q=1)$ \\
\hline
1. & 0.5 & $4.21$ & $7.56$   \\
2. & 1.0 &  $5.4$ & $9.59$   \\
3. & 1.5 &  $5.8$ & $10.16$   \\
4. & 2.0&  $5.93$ & $10.32$  \\
\hline
\hline
S. N. & $q$ & $r_{ps}(\alpha=\mu=\nu=1)$ & $R_s(\mu=\nu=q=1)$ \\
\hline
1. & 0.5 & $3.31$ & $5.76$   \\
2. & 1.0 & $5.4$ & $9.59$   \\
3. & 1.5 & $11.63$ & $20.76$   \\
4. & 2.0 & $24.64$ & $43.68$  \\
\hline
\end{tabular}
\end{table*}

\section{Basic equation for the perturbation of the regular BH  :}
The Klein-Gordon equation around the regular BH   of the massless scalar field
is given by
\begin{eqnarray}
\label{bdelta}
\nabla^2\xi = 0
\end{eqnarray}
\begin{figure}
\begin{center}
{\includegraphics[width=0.55\textwidth]{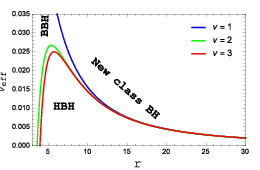}}
\end{center}
\caption{The figure shows the $V_{eff}$ versus $r$. Here, $M =0, \alpha=0.5, q =0.05 $, $\nu =1(\textrm{New class BH}),\nu = 2(\textrm{BBH})$, and
$\nu = 3(\textrm{HBH}).$
\label{4vnu}}
\end{figure}
\begin{figure*}
\begin{tabular}{rl}
\includegraphics[width=0.45\textwidth]{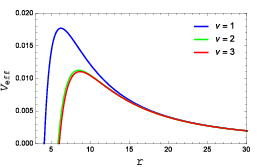}
\includegraphics[width=0.45\textwidth]{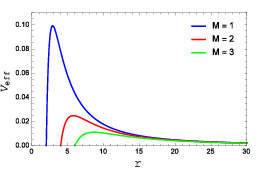}\\
\end{tabular}
\caption{The figure shows the $V_{eff}$ versus $ r$  for varying $\nu$~(left Panel) and $M$~(right panel); the other parameters fixed to $l = 1, \mu=3 $ and $ \alpha= 0.5$.}
\label{4num}
\end{figure*}
\begin{figure*}
\begin{tabular}{rl}
\includegraphics[width=0.45\textwidth]{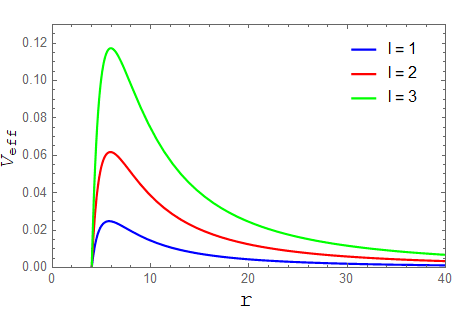}
\includegraphics[width=0.45\textwidth]{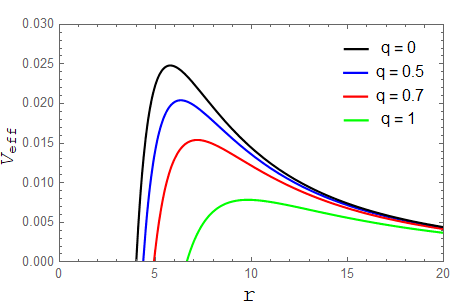}\\
\end{tabular}
\caption{The figure shows the $V_{eff}$ versus $ r$  for varying $l$~(left Panel) and $q$~(right panel); the other parameters fixed to $M = 1, \mu=3 $ and $\alpha= 0.5$.}
\label{4lq}
\end{figure*}
By separation of the variable we have
\begin{eqnarray}
\label{sepa}
\xi = e^{-iwt} X_{l,m}(\theta,\phi)\frac{\eta(r)}{r}
\end{eqnarray}
After simplifying equation $(\ref{bdelta})$ we obtain  a Schr$\ddot{o}$dinger-type  equation which is given by
\begin{eqnarray}
\frac{d^2 \eta(r_*)}{d r_*^2} + \bigg(w^2- V_{eff}(r_*)\bigg)\eta(r_*)=0
\end{eqnarray}
where, $V_{eff}(r_*)$ is given by
\begin{eqnarray}
V_{eff}(r_*) = \bigg[\frac{l (l+1)}{r^2} + \frac{2}{r^3}\bigg(M+ \frac{q^3 r^\mu q^3  (r^\nu + (1- \mu)q^\nu) }{\alpha {(r^\nu +q^\nu)^{(\frac{\mu +\nu}{\nu})}}}\bigg) \bigg]\bigg[ 1-\frac{2 M}{r} - \frac{2 \alpha^{-1}q^{3} r^{\mu-1}}{(r^\nu +q^\nu)^\frac{\mu}{\nu}}\bigg].
\end{eqnarray}
Here,  $w,  X_{l,m}(\theta,\phi)$ and $ r_{*}$ represents the frequency of the wave mode, the spherical
harmonic and the tortoise coordinate, respectively. The effective potential $V_{eff}(r)$ depends on the
parameters $ l, M, \alpha, q, \mu$ and $\nu$.

The comparison of effective potentials among  the Hayward BH~(HBH), Bardeen BH~(BBH) and
new class BH  are shown in Fig.\ref{4vnu}. One can see from Fig.\ref{4vnu} that the
effective potential of new class BH is greatest compared to Bardeen BH and Hayward BH.

In Fig.~\ref{4num}, $V_{eff}$ is plotted against  $r$~(left panel) by varying $\nu$ and
$V_{eff}$ is plotted against $r$~(right panel) by varying $M$. One can see that for both
cases height of $V_{eff}$ is decreasing for increasing of $\nu$ and $M$, respectively.

Fig.~\ref{4lq} represents the plot of $V_{eff}$ versus  $r$ (left panel) by varying $l$
and $V_{eff}$ versus $r$~(right panel) by varying $q$. The height of $V_{eff}$  increases
as $l$ increases and when $q$ increases the height of the $V_{eff}$ decreases.

\section{Massive scalar perturbations:}
In this section, we will see how the massive scalar field decay. It has been observed for
Schwarzschild BH that the massless modes decays faster than massive scalar
field~\cite{Konoplya05}. Hence it is quite interesting to see that if such
characteristics is possible for the above metric~(\ref{met}).\\
Let us take the equation of motion for massive scalar field as
\begin{figure}
\begin{center}
{\includegraphics[width=0.6\textwidth]{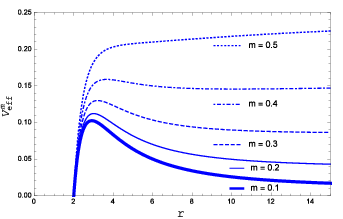}}
\end{center}
\caption{The figure shows the $V^m _{null}$ versus $r$ for varying values of the mass $m$. Here, $M = 1, q = 0.5, \mu =3, \nu = 1, l=1$ and $\alpha = 0.5$.}
\label{veffm}
\end{figure}
\begin{eqnarray}
\label{bdeltanew}
\nabla^2\xi - m^2 \xi = 0.
\end{eqnarray}
After using the separation  of variable similar to the Eq.~(\ref{sepa}), we obtain the
following equation for radial component as
\begin{eqnarray}
\frac{d^2 \eta(r_*)}{d r_*^2} + \bigg(w^2- V^m _{eff}(r_*)\bigg)\eta(r_*)=0,
\end{eqnarray}
and the modified potential $ V^m _{eff}(r_*)$ is given by
\begin{eqnarray}
V^m _{eff}(r_*) = \bigg[\frac{l (l+1)}{r^2} +
\frac{2}{r^3}\bigg(M+ \frac{q^3 r^\mu q^3  (r^\nu + (1- \mu)q^\nu) }{\alpha {(r^\nu +q^\nu)^{(\frac{\mu +\nu}{\nu})}}}\bigg)
+m^2 \bigg]\bigg[ 1-\frac{2 M}{r} - \frac{2 \alpha^{-1}q^{3} r^{\mu-1}}{(r^\nu +q^\nu)^\frac{\mu}{\nu}}\bigg].
\end{eqnarray}
The effective potential $V^m _{eff}(r_*)$ is plotted in the Fig.~\ref{veffm} for different values of the
mass parameter $m$. The Fig.~\ref{veffm} shows that when mass increases the height of the effective
potential also increases. Also the potential terminates to have maximum at the critical values of $m$.
\section{Unstable null geodesics and quasinormal modes of massless scalar field in the eikonal limit}
A quasinormal mode is a solution to the differential equation which has a complex frequency. It
satisfies the boundary condition of purely ``Outgoing" waves, which propagating away from
the boundary, from $-\infty$ to $  +\infty $. WKB method gives the correct approximation of
QNMs at an eikonal limit.

Here, the central equation (wave eqn.) can be taken in the following form
\begin{eqnarray}
\frac{d^{2} \psi}{d x^{2}} + Q(x) \psi = 0,
\end{eqnarray}
where,
\begin{eqnarray}
Q &=& w^2 - V_{eff}(r)\\
\textrm{and}\nonumber\\
V_{eff}(r) &=& \bigg[\frac{l (l+1)}{r^2} + \frac{2}{r^3}\bigg(M+ \frac{q^3 r^\mu q^3  (r^\nu + (1- \mu)q^\nu) }{\alpha {(r^\nu +q^\nu)^{(\frac{\mu +\nu}{\nu})}}}\bigg) \bigg]\bigg[ 1-\frac{2 M}{r} - \frac{2 \alpha^{-1}q^{3} r^{\mu-1}}{(r^\nu +q^\nu)^\frac{\mu}{\nu}}\bigg].
\end{eqnarray}
In case of BH, $ \psi$ represents the radial part of perturbation variable, which assumed to be time dependent.
The coordinate $ x $ is linearly connected  to the ``tortoise" coordinate  $ r_{*} $ ,which  ranges
from $ -\infty $ ( at the horizon) to $ +\infty $ (at the spatial infinity).
The tortoise coordinate  $ r_{*} $ and the radial coordinate $r$ are related by
the following relation
\begin{eqnarray}
\frac{d r}{ d r_{*}} = \bigg(1-\frac{2 M}{r} - \frac{2 \alpha^{-1}q^{3} r^{\mu-1}}{(r^\nu +q^\nu)^\frac{\mu}{\nu}} \bigg).
\end{eqnarray}\\
The function $ Q(x)$, which depends on the angular momentum and the mass of the BH, is
constant at $ x = \pm \infty$.
At $ l\rightarrow\infty $ (i.e.,  in case of eikonal limit), we get
\begin{eqnarray}
\label{psi}
Q_{0} \simeq \omega^{2} - \frac{l^2}{r^{2}} \bigg(1-\frac{2 M}{r} -
\frac{2 \alpha^{-1}q^{3} r^{\nu-1}}{(r^\nu +q^\nu)^\frac{\mu}{\nu}} \bigg),
\end{eqnarray}
where $ l$ represents the angular harmonic index.  \\
From the eqn. (\ref{psi}), the maximum value of $Q_{0}$ at $r = r_{\sigma}$, is given by
\begin{figure*}
\begin{tabular}{rl}
\includegraphics[width=0.45\textwidth]{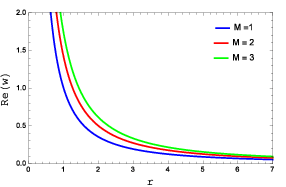}
\includegraphics[width=0.45\textwidth]{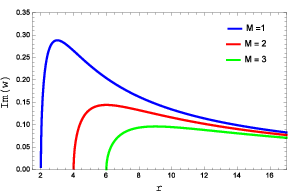}\\
\end{tabular}
\caption{The figure shows QNMs frequency $Re(w)$ versus $ r$ (left Panel) and $Im(w)$ versus $r$ (right panel). Here $q=0.5, \mu = 3, \alpha = 0.5$ and $\nu = 2$.}
\label{5r}
\end{figure*}
\begin{figure*}
\begin{tabular}{rl}
\includegraphics[width=0.45\textwidth]{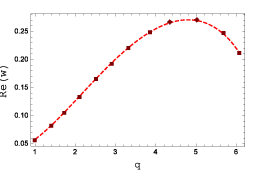}
\includegraphics[width=0.45\textwidth]{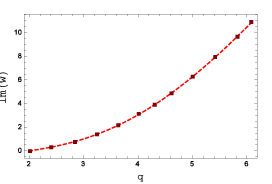}\\
\end{tabular}
\caption{The figure shows QNMs frequency $Re(w)$ versus $ q$ (left Panel) and $Im(w)$ versus $r$ (right panel); the other parameter fixed to $M=1, \mu = 3, \alpha = 0.5$ and $\nu = 2$.}
\label{5q}
\end{figure*}
\begin{figure*}
\begin{tabular}{rl}
\includegraphics[width=0.45\textwidth]{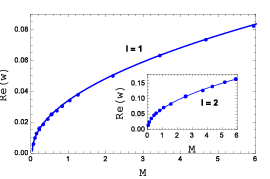}
\includegraphics[width=0.45\textwidth]{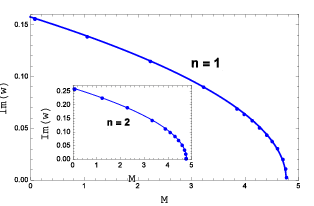}\\
\end{tabular}
\caption{The figure shows QNMs frequency $Re(w)$ versus $ M$ (left Panel) and $Im(w)$ versus $M$ (right panel). Here $q=0.5, \mu = 3, \alpha = 0.5$ and $\nu = 2$.
\label{5m}}
\end{figure*}
\begin{eqnarray}
\bigg(r^\mu_{\sigma} q^3((\mu - 3 ) q^\nu - 3 r^\nu_{\sigma}) + ( r_{\sigma} - 3M) {(r^\nu_{\sigma} +q^\nu)^\frac{\mu +\nu}{\nu}} \alpha\bigg) = 0.
\end{eqnarray}
At $ r = r_o $, the unstable null circular geodesic can be calculated by using
$  \dot{r}^2 = (\dot{r}^2)' = 0$, we have
\begin{eqnarray}
\bigg(r^\mu_{o} q^3((\mu - 3 ) q^\nu - 3 r^\nu_{o}) + ( r_{o} - 3M) {(r^\nu_{o} +q^\nu)^\frac{\mu +\nu}{\nu}} \alpha\bigg) = 0.
\end{eqnarray}
At the point $  r_{\sigma} = r_{o},$ the maximum value of $ Q_{0}$ and
the null circular geodesics  are coincident, then the QNM leads to the
following form
\begin{eqnarray}
\label{in}
\frac{Q_{0} (r_{\sigma})}{\sqrt{2 Q_{0}^{''}(r_{\sigma})}} = i ( n + 1/2),
\end{eqnarray}
where  $ Q_{0}^{''} \equiv \frac{d^{2} Q_{0} }{d r_{*}^{2}}$ and the Eq.~(\ref{in})
is calculated at an extremum of $ Q_{0}$~(that is, $ \frac{d Q_{0}}{d r_{*}} = 0$ at $r_0$).  \\

Following the formula  that has been derived by Cardoso et al.~\cite{Cardoso09} as
\begin{eqnarray}
\label{qnm}
\omega_{QNM} = l\Omega_o -i ( n + 1/2)\lambda_o,
\end{eqnarray}
where $n$ represents the overtone number, $\Omega_o$ is  the angular frequency
measured by the asymptotic  observers  and $\lambda_0$ is the coordinate
Lyapunov exponent of null-circular geodesics.  Angular frequency~($\Omega_o$)
and Lyapunov exponent ($\lambda_0$) for null-circular geodesics  
are studied in Appendix.
In case of eikonal approximation~(i.e., in the large-$l$ limit) the QNM frequency for
BH ~\cite{Mondal/20}can be represented by the following two parameters:
\begin{eqnarray}
\label{qnm}
&&\omega_{QNM} = l\sqrt{\frac{M +q^3 r^\mu_{o} q^3 \alpha^{-1} (r^\mu_{o} + (1- \mu)q^\nu){(r^\nu_{o} +q^\nu)^{-(\frac{\mu +\nu}{\nu})}} }{r^3_{o}}} -i( n + 1/2)\times \nonumber\\
&&  \frac{\sqrt{( (r_{o} - 2 M)(r^\nu_{o} +q^\nu)^\frac{\mu }{\nu} \alpha - 2 q^3 r^\mu_{o})(   r_{o}  \alpha (r^\nu_{o} +q^\nu)^{2+\frac{\mu }{\nu}}   + \mu  q^{(3+ \nu) } r^\mu_{o}( 3q^\nu ( \nu - 3) - r^\nu_{o}(\nu +3))}}{r^2_{o} \alpha(r^\nu_{o} +q^\nu)^\frac{\mu +\nu}{\nu} }.
\end{eqnarray}
Which is the key results of our manuscript and
the importance  of the equation~(\ref{qnm}) is that in case of  eikonal limit, the real
and complex parts of the QNMs for the spherically symmetric, asymptotically flat
space-time are stated by the frequency and instability time scale  for the case of
unstable null circular geodesics.

Fig.~\ref{5r} represents the plot of $Re(w)$ versus  $q$~(left panel) and
$Im(w)$ versus $q$ (right panel) by varying $M$. $Re(w)$ decreases as $r$
increases and when $M$ increases, the height of the $Im(w)$ decreases.

In Fig.\ref{5q}, $Re(w)$ is graphed versus $q$ (left panel) and $Im(w)$ is
graphed versus $q$ (right panel). One can see that $Re(w)$ is increased for the
initial value of $q$. However, it seems $Re(w)$ decrease for large value of
$q$ as evident from Fig.\ref{5q}. But the $Im(w)$ is always increasing  with
increasing the value of $q$.

In Fig.~\ref{5m}, $Re(w)$ is graphed versus $M$~(left panel) for both $l=1$
and $l = 2$ . $Im(w)$ is graphed versus $M$~(right panel) for both $n=1$ and
$ n= 2$. $Re(w)$ increases with $M$ and $Im(w)$ is decreased when $M$ increased.
Interestingly, when $M$ increases, there is a critical point where $Im(w)$
approaches to zero leading to purely real modes.

\section{\label{} Time-like Geodesics of Regular BHs }
\subsection{The Effective Potential}
\begin{figure}
\centering
\includegraphics[width=7cm,height=5cm,angle=0]{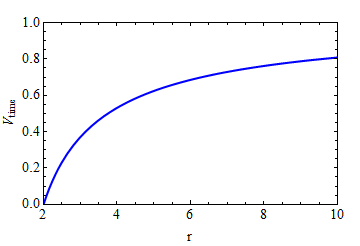}
\includegraphics[width=7cm,height=6cm,angle=0]{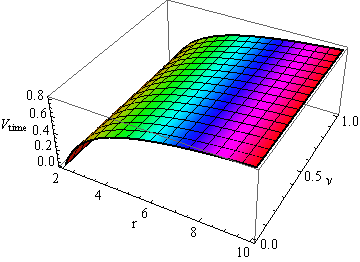}
\includegraphics[width=7cm,height=5cm,angle=0]{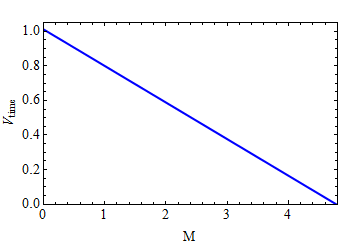}
\includegraphics[width=7cm,height=6cm,angle=0]{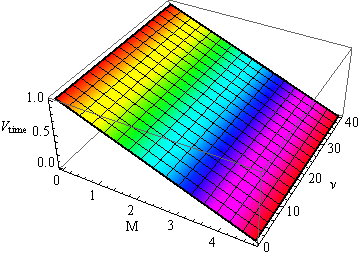}
\includegraphics[width=7cm,height=5cm,angle=0]{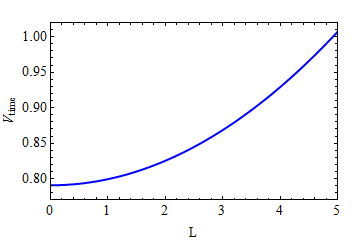}
\includegraphics[width=7cm,height=6cm,angle=0]{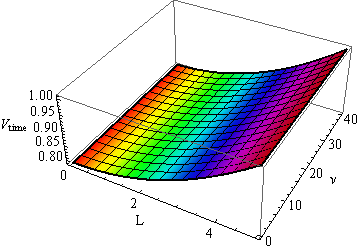}
\includegraphics[width=7cm,height=5cm,angle=0]{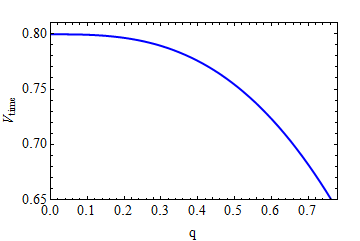}
\includegraphics[width=7cm,height=6cm,angle=0]{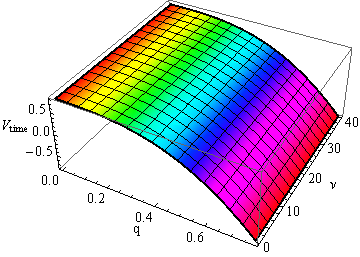}
\caption{Plot of $V_{time}$ versus  $r$ (first panel),
$M$ (second panel), $L$ (third panel) and $q$ (fourth panel) respectively. We have set ${\mu}=3$.
\label{vtime}}
\end{figure}
For time-like circular geodesics~($\delta = -1 $) the radial Eq.(\ref{dotr}) reduces to
\begin{eqnarray}
\label{timerad}
\dot{r}^{2} = E^{2} - V_{time}= E^{2}-  \bigg( 1 + \frac{L^{2}}{r^{2}}\bigg)
\bigg(1-\frac{2 M}{r} - \frac{2 \alpha^{-1}q^{3} r^{\mu-1}}{(r^\nu +q^\nu)^\frac{\mu}{\nu}} \bigg),
\end{eqnarray}
$V_{time}$ denoted the effective potential for time-like geodesics which is given by
\begin{eqnarray}
 V_{time}= \bigg( 1 + \frac{L^{2}}{r^{2}}\bigg)\bigg(1-\frac{2 M}{r}
 - \frac{2 \alpha^{-1}q^{3} r^{\mu-1}}{(r^\nu +q^\nu)^\frac{\mu}{\nu}} \bigg).
\end{eqnarray}
The geodesic motion of neutral test particles can be analyzed by using the effective
potential diagram which is graphically shown in Fig.~\ref{vtime}.

Analogously, the effective potential with zero angular momentum geodesics is
\begin{eqnarray}
V_{time}= \bigg(1-\frac{2 M}{r} - \frac{2 \alpha^{-1}q^{3} r^{\mu-1}}{(r^\nu +q^\nu)^\frac{\mu}{\nu}} \bigg),
\end{eqnarray}
To calculate the circular geodesics motion of the test particle, we will use the condition $\dot{r}=0$ and $\dot{r}'=0$ at $r=r_c$.
From the Eq.~(\ref{timerad}) we get
\begin{eqnarray}
V_{time} =0,
\end{eqnarray}
and
\begin{eqnarray}
\frac{d V_{time}}{d r} =0.
\end{eqnarray}
Hence the energy and angular momenta per unit mass of the test particle are as follows:

\begin{eqnarray}
E_{c}^{2}& =& \frac{(r^\nu_{c} +q^\nu)^{1 - \frac{\mu}{\nu}}
\bigg((r - 2M) (r^\nu_{c} +q^\nu)^{\frac{\mu}{\nu}}\alpha -2 q^3 r^\mu_{c}\bigg)^{2}}
{r_{c}  \alpha \bigg(r^\mu_{c} q^3((\mu - 3 ) q^\nu - 3 r^\nu_{c})
+ ( r_{c} - 3M) {(r^\nu_{c} +q^\nu)^\frac{\mu +\nu}{\nu}} \alpha\bigg)},
\end{eqnarray}
\begin{figure}
\centering
\includegraphics[width=7cm,height=5cm,angle=0]{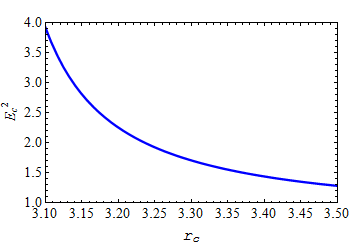}
\includegraphics[width=7cm,height=6cm,angle=0]{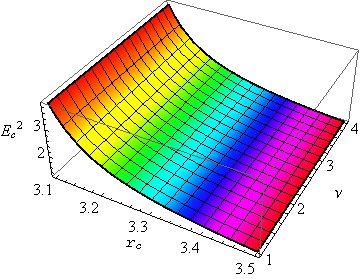}
\caption{Plot of $E^2_c$ versus $r_{c}$ for $M = 2$, $q = 0.05$, $\mu = 3$ and $\alpha = 0.5$. The left panel is plotted for $\nu = 1$ and  right panel is  plotted for values of
$\nu$ in the range $0\leq \nu \leq 4$.
\label{Evr}}
\end{figure}
\begin{figure}
\centering
\includegraphics[width=7cm,height=5cm,angle=0]{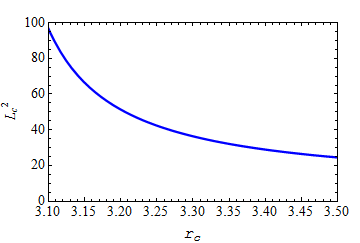}
\includegraphics[width=7cm,height=6cm,angle=0]{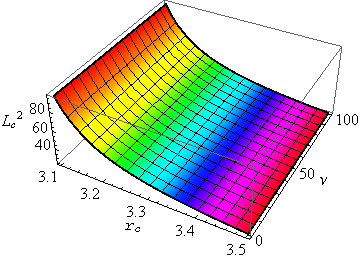}
\caption{Plot of $L^2_{c}$ versus $r_{c}$ for $M = 2$, $q = 0.05$, $\mu = 3$
and $\alpha = 0.5$. The left panel is graphed for $\nu = 1$
and  right panel is graphed for values of $\nu$ in the range $0\leq \nu \leq 100$.
\label{Lvr}}
\end{figure}
and
\begin{eqnarray}
L_{c}^{2} &=& \frac{r^2_{c} \bigg(q^3 r^\mu_{c} (r^\nu_{c} - (\mu - 1)q^\nu)
+ M(r^\nu_{c} +q^\nu)^\frac{\mu +\nu}{\nu}\alpha \bigg)}{\bigg(r^\mu_{c} q^3((\mu - 3 ) q^\nu
- 3 r^\nu_{c}) + ( r_{c} - 3M) {(r^\nu_{c} +q^\nu)^\frac{\mu +\nu}{\nu}} \alpha\bigg)}.
\end{eqnarray}
The 2D and 3D diagram of variation of energy $(E_c)$   and angular momentum $(L_c)$ along
time-like circular geodesics  could be seen from Figs.~\ref{Evr} and \ref{Lvr}, respectively.\\
To exist the circular motion of test particle, the energy and angular momentum must be
real and finite.

Therefore we require,\\

$\bigg(r^\mu q^3((\mu - 3 ) q^\nu - 3 r^\nu) + ( r - 3M) {(r^\nu +q^\nu)^\frac{\mu +\nu}{\nu}} \alpha\bigg) > 0$,\\

and\\

$ \bigg(q^3 r^\mu (r^\nu - (\mu - 1)q^\nu) + M(r^\nu +q^\nu)^\frac{\mu +\nu}{\nu}\alpha \bigg) > 0$.\\

The equality with  limit indicates a circular orbit with diverging energy per unit rest mass, that is,
a photon orbit. This photon orbit is the inner most boundary of the time-like circular orbits for
particles.\\

The orbital velocity is given by
\begin{eqnarray}
\Omega_{c} =\frac{\dot{\phi}}{\dot{t}} = \sqrt{\frac{M +q^3 r^\mu_{c} q^3 \alpha^{-1} (r^\nu_{c}
+ (1- \mu)q^\nu){(r^\nu_{c} +q^\nu)^{(-\frac{\mu +\nu}{\nu})}} }{r^3_{c}}}.
\end{eqnarray}

\subsection{Marginally bound circular orbit~(MBCO):}
The equation of MBCO   looks like
\begin{eqnarray}
4M\alpha^2 r_c (r_c-M)(r^\nu_{c} +q^\nu)^{(\frac{\mu +\nu}{\nu})} &+&4 \alpha (r_c-2M)r_c^{\mu} q^3 (r^\nu_{c} +q^\nu)^{\frac{\mu}{\nu}}\nonumber\\
&-& 4q^6 r_c^{2\mu} (r^\nu_{c} +q^\nu)^{(\frac{\nu -\mu}{\nu})}+ r_{c}^{\mu+1} q^3 \alpha((\mu - 3 ) q^\nu - 3 r^\nu_{c})=0
\end{eqnarray}
Let $r_c = r_{mb}$ be the solution of the above equation which gives the radius of MBCO close to the Regular BH.
\subsection{Equation of ISCO: }
The equation of innermost stable circular orbit~(ISCO) can be obtained from the second derivative of the
effective potential~$(V_{time})$ of the time like geodesics, that is,
\begin{eqnarray}
\frac{d^2 V_{time}}{d r^2} =0.
\end{eqnarray}
\begin{figure}[h!]
\centering
\includegraphics[width=7cm,height=5cm,angle=0]{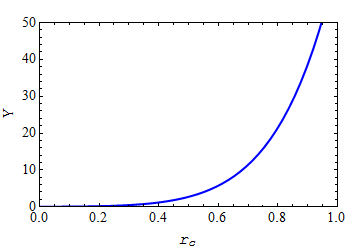}
\includegraphics[width=7cm,height=6cm,angle=0]{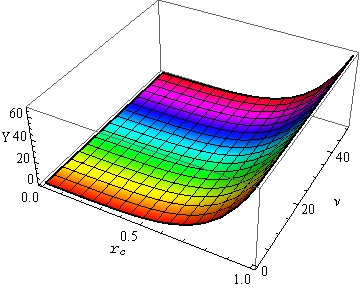}
\includegraphics[width=7cm,height=5cm,angle=0]{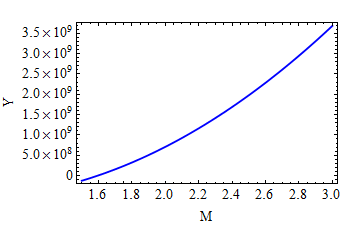}
\includegraphics[width=7cm,height=6cm,angle=0]{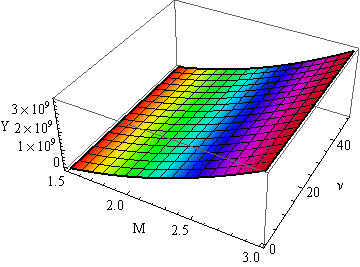}
\includegraphics[width=7cm,height=5cm,angle=0]{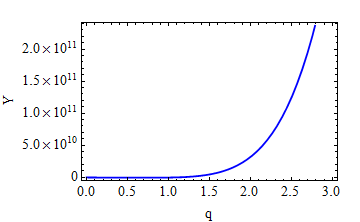}
\includegraphics[width=7cm,height=6cm,angle=0]{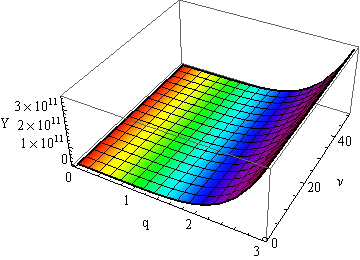}
\includegraphics[width=7cm,height=5cm,angle=0]{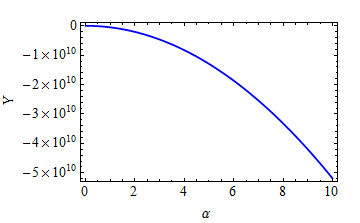}
\includegraphics[width=7cm,height=6cm,angle=0]{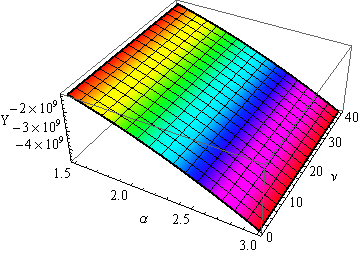}
\caption{The figure depicts the variation of
$Y=2 r^{2\mu}_{c} q^6 s +r^\mu_{c} q^3 \alpha (r^\nu_{c}
+q^\nu)^{\frac{\mu}{\nu}}( t+w)- M (r_{c} - 6 M)(r^\nu_{c}
+q^\nu)^{2(\frac{\mu +\nu}{\nu})}\alpha^2 $ versus $r_{c}$~(first panel),
$M$~(second panel), $q$~(third panel) and $\alpha$~(fourth panel)
respectively. We have set ${\mu}=3$.
\label{isco}}
\end{figure}
Thus the equation of ISCO is
\begin{eqnarray}
\label{isco}
2 r^{2\mu}_{c} q^6 s +r^\mu_{c} q^3 \alpha (r^\nu_{c} +q^\nu)^{\frac{\mu}{\nu}}(t+w)- M (r_{c} - 6 M)
(r^\nu_{c} +q^\nu)^{2(\frac{\mu +\nu}{\nu})}\alpha^2  = 0,
\end{eqnarray}
where
\begin{eqnarray}
s &=& (\mu -3)(\mu -1) q^{2\nu} + (\mu (\nu - 4)+6)r^{\nu}_{c} q^{\nu} + 3 r^{2 \nu}_{c}, \nonumber\\
t &=& r^{2 \nu}_{c}(12 M-r_{c})-  q^{2 \nu}((\mu^2 + 4 \mu -6)2 M +( 1 - \mu^2)r_{c}), \nonumber\\
w &=&  r^{ \nu}_{c} q^{ \nu}((\mu \nu - 4 \mu +12)2M - (2 + \mu \nu)r_{c}).
\end{eqnarray}
Let $r_c = r_{ISCO}$ be the smallest real root of the Eq.~(\ref{isco}) which gives the
radius of the ISCO of the regular BH. When $q\rightarrow 0$, we obtain the
radius of ISCO for Schwarzschild BH which occurs at $r_{ISCO} = 6M$.
\section{CONCLUSION AND FUTURE WORK:}
We investigated the null geodesics of regular BHs. A complete geodesic
study has been made both for time-like geodesics and null geodesics. Studies of
test particle both for photon and massive particles as an interesting approach
to understand the strong gravity around the BH spacetime. As an application of
null geodesics, we derived the radius of photon sphere and gravitational bending
of light. We also studied the shadow of the BH spacetime. Moreover, we showed the
relation between radius of photon sphere~$(r_{ps})$ and the shadow observed by a
distance observer.\\

Using null-circular geodesics, we evaluated the celestial coordinates~$(\beta,\gamma)$ and
the radius $R_s$ of the regular BH shadow and presented it graphically. The effect of dimensionless
constant $\mu$ of the BH, the free integration constant $q$ and the other parameter~(like $\alpha$
and $\nu$) on the radius of shadow are studied in detail. In particular, the radius of BH shadow is
increased with increasing value of $\nu$ and $q$ while the radius is decreased with increasing value
of the parameters~$\alpha$ and $\mu$ respectively.\\

Moreover, we discussed the effect of various parameters on the radius of shadow
$R_s$. Also we computed the angle of deflection for the photons as a physical
application of null-circular geodesics.  We determined the relation between null geodesics
and quasinormal modes frequency in the eikonal approximation by computing the
Lyapunov exponent. It was also shown that~(in the eikonal limit) the quasinormal
modes~(QNMs) of BHs are governed by the parameter of
null-circular geodesics. The real part of QNMs frequency determined the
angular frequency whereas the imaginary part determined the instability
time scale of the circular orbit.\\

Quite apart from the circular geodesics analogy, we determined the relation between unstable
null-circular geodesics and QNMs, which is quite general and being valid for large limit for
any spherically symmetric, static and asymptotically flat space-time. We showed that the
QNMs in the eikonal limit of the BH are defined by the parameter of the null-circular
geodesics. The real part of the complex QNMs frequencies is related to the angular
velocity in the unstable null-circular geodesics and the imaginary part is determined
by the instability timescale of the orbit. More specifically, we showed that the
Lyapunov exponent  and the angular velocity $(\Omega_o)$ at the unstable null
circular geodesics, governing the instability timescale for the orbit admit
with an analytic WKB approximation of QNMs. \\

Furthermore,  we examined the massless scalar perturbations
and analyzed the effective potential graphically. We also studied the massive
scalar perturbations around the BH spacetime. As an application of time-like
geodesics we computed the ISCO and MBCO  of the regular BHs 
which was closely related to the BH accretion disk theory.
In the appendix, we showed the relation between
angular frequency and Lyapunov exponent for null-circular geodesics.
\section*{Acknowledgments}
FR would like to thank the authorities of the Inter-University Centre for Astronomy and Astrophysics, Pune, India for providing research facilities.   SI is also thankful to CSIR for financial support. This work is a part of the project  submitted in DST-SERB, Govt. of India. We are thankful to the honorable referee for his/her comments and constructive suggestions  \\


\section*{Appendix : Angular frequency and Lyapunov exponent for null-circular geodesics}
Angular frequency  of regular BH  is given by
\begin{eqnarray}
 \Omega_{o} =\frac{\dot{\phi}}{\dot{t}}= \sqrt{\frac{f'_o}{2 r_o}}= \frac{f^{\frac{1}{2}}_o}{r_o}.
\end{eqnarray}
Using the equation (\ref{fr})we obtain the angular frequency as follows
\begin{eqnarray}
 \Omega_{o}= \sqrt{\frac{M +q^3 r^\mu_{o} q^3 \alpha^{-1} (r^\nu_{o} + (1- \mu)q^\nu){(r^\nu_{o} +q^\nu)^{(-\frac{\mu +\nu}{\nu})}} }{r^3_{o}}} .
\end{eqnarray}
The Lyapunov exponent can be defined in terms of the second derivative of the effective potential for radial motion $V_{null}$ of null-circular geodesics as
\begin{eqnarray}
\label{lao}
\lambda_{o} = \sqrt{\frac{(V_{null})''}{2\dot{t}^{2}}}.
\end{eqnarray}
\begin{figure}[h!]
\centering
\includegraphics[width=7cm,height=5cm,angle=0]{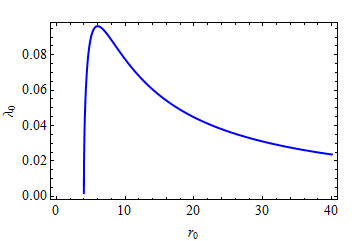}
\includegraphics[width=7cm,height=6cm,angle=0]{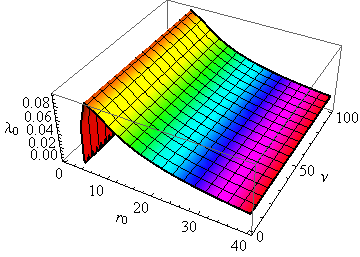}
\caption{Plot of $\lambda_{0}$ versus $r_{0}$ for $M = 2$, $q = 0.05$, $\mu = 3$ and $\alpha = 0.5$. The left panel is graphed for $\nu = 1$ and  right panel is graphed for values of $\nu$ in the range $0\leq \nu \leq 100$.
\label{lamdao}}
\end{figure}
Using the equation (\ref{lao}) the Lyapunov exponent for null circular geodesics is given by
\begin{eqnarray}
\lambda_{o} = \frac{\sqrt{\bigg( (r_{o} - 2 M)(r^\nu_{o} +q^\nu)^\frac{\mu }{\nu} \alpha - 2 q^3 r^\mu_{o} \bigg)\bigg(   r_{o}  \alpha (r^\nu_{o} +q^\nu)^{2+\frac{\mu }{\nu}}   + \mu  q^{(3+ \nu) } r^\mu_{o}( 3q^\nu ( \nu - 3) - r^\nu_{o}(\nu +3))  \bigg)}}{r^2_{o} \alpha(r^\nu_{o} +q^\nu)^\frac{\mu +\nu}{\nu} }.
\end{eqnarray}
We have plotted the graph $\lambda_{o}$ against $r_o$ in the Fig.\ref{lamdao}.
From the Fig.\ref{lamdao} it is clear that $\lambda_{o}$ increases for the initial
values of $r_o$ and after that it starts to decrease for large value of $r_o$ .



\begin{thebibliography}{99}

\bibitem{Zhang/1997} S. N. Zhang, W. Cui and W. Chen, ``Black Hole Spin in X-Ray Binaries: Observational Consequences'' {\it Astrophys. J.} {\bf 482}, L155 (1997) [arXiv:astro-ph/9704072].

\bibitem{Kokkotas/1999} K. D. Kokkotas and B. G. Schmidt, ``Quasi-Normal Modes of Stars and Black Holes'' {\it Living Rev. Rel.} \textbf{2}, 2 (1999).

\bibitem{Nollert/1999} H. P. Nollert, ``Quasinormal modes: the characteristic `sound' of black holes and neutron stars'' {\it Class. Quant. Grav.} {\bf 16}, R159 (1999).

\bibitem{Mondal/2021} M. Mondal, F. Rahaman, K. N Singh, ``Lyapunov exponent, ISCO and Kolmogorov?Senai entropy for Kerr?Kiselev black hole'' {\it Eur. Phys. J. C} {\bf 81}, 84 (2021).

\bibitem{Cornish03} N. J. Cornish and J. J Levin, ``Lyapunov timescales and black hole binaries'' {\it Class. Quant. Grav.} {\bf 20}, 1649 (2003) [arXiv:gr-qc/0304056].

\bibitem{Berti/2009} E. Berti, V. Cardoso A. O. Starinets, ``Quasinormal modes of black holes and black branes'' {\it Class. Quant. Grav.} {\bf 26}, 163001 (2009).

\bibitem{Konoplya11} R. A. Konoplya, ``Quasinormal modes of black holes: From astrophysics to string theory'' {\it Rev Mod. Phys} {\bf 83}, 793 (2011).

\bibitem{Mashhoon/1985} B. Mashhoon, ``Stability of charged rotating black holes in the eikonal approximation'' {\it Phys. Rev. D} {\bf 31}, 290 (1985).

\bibitem{Berti/2005} E. Berti and K. D. Kokkotas, ``Quasinormal modes of Kerr-Newman black holes: coupling of electromagnetic and gravitational perturbations'' {\it Phys. Rev. D} {\bf 71}, 124008 (2005).

\bibitem{Pretorius07} F. Pretorius and D. Khurana, ``Black hole mergers and unstable circular orbits'' {\it Class. Quant. Grav.} {\bf 24}, S83 (2007).

\bibitem{Steklain/2009} A. F. Steklain, P. S. Letelier, ``Stability of orbits around a spinning body in a pseudo-Newtonian Hill problem'' {\it Physics Letters A} \textbf{373}, 188 (2009).

\bibitem{Schnittman/2001} J. D. Schnittman, F. A. Rasio, ``Ruling Out Chaos in Compact Binary Systems'' {\it Phys. Rev. Lett.} {\bf 87}, 121101 (2001).

\bibitem{Barrow/1981} J. D. Barrow, ``Chaos in the Einstein Equations'' {\it Phys. Rev. Lett.} {\bf 46}, 963 (1981).

\bibitem{Burd/1994} A. Burd and A. Coley, ``Deterministic Chaos in General Relativity eds. D. Hobill", {\it Plenum Press}, New York, (1994).

\bibitem{Moni19} T. Manna, F.Rahaman, M. Mondal, ``Solar system tests in Rastall gravity", {\it Modern Physics Letter  A} 2050034, (2019).

\bibitem{Semerak/1999} O. Semerak and V. Karas, ``Pseudo-Newtonian models of a rotating black hole field '' {\it Astron. Astrophys.} {\bf 343}, 325 (1999).

\bibitem{Das/2020}  S. Das, N. Sarkar,M. Mondal, F. Rahaman, ``A new model for dark matter fluid sphere", Mod. Phys. Lett. A 35,
2050280 (2020)

\bibitem{Dettmann/1994} C. P. Dettmann, N. E. Frankel, N.J. Cornish, ``Fractal basins and chaotic trajectories in multi-black-hole spacetimes'' {\it Phys. Rev. D} {\bf 50}, R618 (1994).

\bibitem{Cornish/1997} N. J. Cornish and J. J. Levin, `` The mixmaster universe is chaotic'' {\it Phys. Rev. Lett.} {\bf 78}, 998 (1997).

\bibitem{Prasobha/2014} C. B. Prasobha, V. C. Kuriakose, ``Quasinormal modes of Lovelock black holes" {\it Eur. Phys. J. C} {\bf 74}, 3136 (2014).

\bibitem{Motl/2003} L. Motl, A. Neitzke, ``Asymptotic black hole quasinormal frequencies" {\it Adv. Theor. Math. Phys.} {\bf7}, 307 (2003).

\bibitem{Dreyer/2003} O. Dreyer,``Quasinormal modes, the area spectrum, and black hole entropy" {\it Phys. Rev. Lett.} {\bf 90}, 081301 (2003).

\bibitem{Fernando/2014} S. Fernando, T. Clark, ``Black holes in massive gravity: quasi-normal modes of scalar perturbations'' {\it Gen. Relativ. Gravit.} {\bf 46}, 1834 (2014).

\bibitem{Hendi/2020} S. H. Hendi, A. Nemati, ``Thermodynamics shadow and quasinormal modes of black holes
in five-dimensional Yang-Mills massive gravity'', arXiv:1912.06824v2 (2020)

\bibitem{Ghosh/2014} S. G. Ghosh, P. Sheoran, M. Amir, ``Rotating Ayon-Beato-Garcia black hole as a particle accelerator'' {\it Phys. Rev. D} {\bf 90}, 103006 (2014).

\bibitem{Ghosh/2015} S. G. Ghosh, M. Amir, ``Horizon structure of rotating Bardeen black hole and particle acceleration'' {\it Euro. Phys. J. C} {\bf 7}, 553 (2015).

\bibitem{Amir/2016} M. Amir, F. Ahmed, S. G. Ghosh, ``Collision of two general particles around a rotating regular Hayward?s black holes'' {\it Euro. Phys. J. C}, {\bf 76} 532 (2016).

\bibitem{Toshmatov/2015} B. Toshmatov, A. Abdujabbarov, Z. Stuchik, B. Ahmedov, ``Quasinormal modes of test fields around regular black holes'' {\it Phys. Rev. D} {\bf 91}, 083008 (2015). 

\bibitem{Fan/2016} Z-Y Fan, and X Wang,``Construction of regular black holes in general relativity" {\it Phys. Rev.D} {\bf 94}, 124027 (2016).

\bibitem{Bardeen74}  J.M. Bardeen, in Conference Proceeding of GR5, Tbilisi,USSR, p. 174 (1968).

\bibitem{Hayward06} S. A. Hayward, ``Formation and evaporation of non-singular black hole", \textit{Phys. Rev.Lett.}, \textbf{96}, 031103 (2006).

\bibitem{Chandrasekhar83} S. Chandrasekhar, ``The Mathematical Theory of Black Holes'', {\it Oxford University Press}, New York, (1983).

\bibitem{Kimet/20} K.Jusufi, ``Quasinormal modes of black holes surrounded by dark matter
and their connection with the shadow radius". \textit{Phys. Rev D.},\textbf{ B 101}, 084055 (2020).

\bibitem{Das20} A. Das, A. Saha and S. Gangopadhyay,``Shadow of charged black holes in Gauss Bonnet gravity'' {\it Eur. Phys. J. C} {\bf 80}, 180 (2020).

\bibitem{chiba17} T. Chiba,  M. Kimura, ``A note on geodesics in the Hayward metric'' {\it Prog. Theor. Exp. Phys.} {\bf 043E01} (2017).

\bibitem{Cardoso09} V. Cardoso, A. S. Miranda, E. Berti, H. Witek, and V. T. Zanchin,``Geodesic stability, Lyapunov exponents and quasinormal modes", \textit{Phys. Rev.}, \textbf{D 79} 064016 (2009).

\bibitem{Mondal/20} M. Mondal, P. Pradhan, F. Rahamann, I. Karar, arXiv.org/abs/2008.11022(2020).

\bibitem{Konoplya05} R. A. Konoplya, A.V. Zhidenko, ``Decay of massive scalar field in a Schwarzschild background". \textit{Phys. Lett.},\textbf{ B 609}, 377 (2005).



\end{thebibliography}
\end{document}